\documentclass[aps,pra,twocolumn,showpacs,landscape=false]{revtex4-1}

\usepackage[dvips]{graphicx}
\usepackage{caption}
\usepackage{subcaption}

\usepackage{amsmath,amssymb}
\usepackage{amsfonts}
\usepackage{color}

\usepackage[normalem]{ulem} 
\usepackage{placeins}
\usepackage{mdframed} 
\usepackage{soul} 
\usepackage{comment} 
\usepackage{MnSymbol}
\usepackage{tikz}
\usepackage{pgfplots}
\usepackage{cleveref}
\usepackage{float}
\usepackage{soul}
\usepackage{lipsum}
\usepackage{readarray}
\usepackage{textgreek}
\usepackage{braket}

\usepgfplotslibrary{smithchart}

\usetikzlibrary{positioning}

\newcommand\Csref[1]{\textbf{\subref{#1}},}

\newcommand\dB{\ensuremath{\,\mathrm{dB}}}
\newcommand\dBm{\ensuremath{\,\mathrm{dBm}}}

\newcommand\ie{\textit{i.e.}}

\newcounter{tikznumber}
\newcommand\abs[2][]{\ensuremath{\left|#2\right|^{#1}}}

\newcommand\pln[2][]{\ensuremath{\ln^{#1}\!\left(#2\right)}}

\newcommand\J[1][]{\ensuremath{_{\mathrm{J}#1}}}

\newcommand\crit[1][]{\ensuremath{_{\mathrm{c}#1}}}

\newcommand\pump[1][]{\ensuremath{_{\mathrm{p}#1}}} 
\newcommand\signal[1][]{\ensuremath{_{\mathrm{s}#1}}} 
\newcommand\res{\ensuremath{_\mathrm{res}}}

\newcommand\signed[1]{\ifnum #1>0 {+#1} \else{\ifnum #1<0 {#1} \fi} \fi}
\newcommand\coeff[1]{
    \ifnum #1=1 {+} \else{
        \ifnum #1=-1 {-} \else{
            \ifnum #1>0 {+#1} \else{
                \ifnum #1<0 {#1} \fi
            } \fi
        } \fi
    } \fi
}

\input{settings/tikzicons.tex}

\Crefname{equation}{Eq.}{Eqs.}
\Crefname{figure}{Fig.}{Figs.}
\Crefname{tabular}{Tab.}{Tabs.}
\setstcolor{red}
\pgfplotsset{compat=1.3,tick scale binop=\times}

\definecolor{mycolor1}{rgb}{0.00000,0.44700,0.74100}%
\definecolor{mycolor2}{rgb}{0.85000,0.32500,0.09800}%
\definecolor{mycolor3}{rgb}{0.92900,0.69400,0.12500}%
\definecolor{mycolor4}{rgb}{0.49400,0.18400,0.55600}%
\definecolor{mycolor5}{rgb}{0.46600,0.67400,0.18800}%
\definecolor{mycolor6}{rgb}{0.30100,0.74500,0.93300}%
\definecolor{mycolor7}{rgb}{0.63500,0.07800,0.18400}%
\definecolor{BLUE}{rgb}{0, 0, 1}

\makeatletter
\AtBeginDocument{\let\LS@rot\@undefined}
\makeatother

\begin{document}

\title{A small footprint travelling-wave parametric amplifier with a high Signal-to-Noise Ratio improvement in a wide band}
\author{Hampus Renberg Nilsson}
\email{Hampus.Renberg.Nilsson@chalmers.se}
\author{Liangyu Chen}
\author{Giovanna Tancredi} 
\author{Robert Rehammar} 
\author{Daryoush Shiri} 
\author{Filip Nilsson} 
\author{Amr Osman} 
\author{Vitaly Shumeiko} 
\author{Per Delsing} 
\address{Department of Microtechnology and Nanoscience - MC2, Chalmers University of Technology,
S-412 96 G\"oteborg, Sweden.}
\date\today

\begin{abstract}

We characterise a small footprint travelling-wave parametric amplifier (TWPA).
The TWPA is built with magnetically flux-tunable superconducting nonlinear asymmetric inductive elements (SNAILs) and parallel-plate capacitors.
It implements three-wave mixing (3WM) with resonant phase matching (RPM),
a small cutoff frequency for high gain per unitcell
and impedance matching networks for large bandwidth impedance matching.
The device has 200 unitcells and a physical footprint of only 1.1\,mm\(^2\),
yet demonstrates an average parametric gain of 19\,dB over a 3\,GHz bandwidth,
an average effective signal-to-noise ratio improvement of 10\,dB
and a clear speedup of qubit readout time.

\end{abstract}

\maketitle 

\section{Introduction}

There is a high interest in building large-scale superconducting quantum computers.
To achieve high fidelity readout, quantum-limited amplifiers \cite{Caves1982} with a high gain are desired.
They are typically built as superconducting nonlinear oscillators or transmission lines
and have demonstrated high gain with near-quantum-limited noise performance \cite{Yamamoto2008,Bergeal2010,Roch2012,Roy2016,AumentadoRev}
and have become an essential part of the circuit Quantum Electrodynamics (cQED) \cite{Girvin2008} toolbox.
While Josephson parametric amplifiers (JPAs) have shown high gain and near-quantum-limited noise performance \cite{Kanter1971,Feldman1975,Yurke1988,Yurke1989,Olsson1988, Beltran2008,Roy2015,Simoen2015},
and achieved high fidelity readout of a single qubit,
they typically only have a high gain over a narrow band,
making them unsuitable for multiplexed readout in large-scale quantum computers.

The capability of quantum-limited noise performance and a high gain in a wide band is provided by
the travelling-wave parametric amplifier (TWPA)~\cite{Cullen1958,Suhl1958,Tien1958}.
The basic principle for amplification of the TWPA is based on frequency-mixing between
a weak signal and a strong wave, the pump,
when they copropagate through a nonlinear medium.
When a phase-matching condition between the pump, the signal and the idler is fulfilled,
this can result in exponential spatial growth of the signal amplitude.

Different TWPAs can be distinguished in several ways:
i) what nonlinear element they use,
typically superconductor kinetic inductance \cite{LeDuc2012,Pappas2014,Pappas2016,Gao2017,Erickson2017,Katz2020,GaoPRXQ2021, Duti2021}
or the inductance of Josephson junctions \cite{Siddiqi2013,Obrien2014,Macklin2015,Martinis2015,Bell2015,Planat2020,Ranadive2022,Zorin2016,Zorin2019,Sivak2019,Mukhanov2019,Aalto2021},
ii) whether they implement three-wave mixing (3WM) or four-wave mixing (4WM),
and iii) what kind of dispersion engineering is used,
typically either resonant phase matching~(RPM)~\cite{Obrien2014,Macklin2015,MyTWPA3WM},
periodic modulation~(PM)~\cite{Katz2020,GaoPRXQ2021,AnitasArticle,Gaydamachenko2022}
or reversed Kerr effect~\cite{Ranadive2022}.

An important property of any TWPA is the loss,
and its impact on the noise performance.
While a lossless TWPA would, in theory, have quantum-limited noise performance,
a real TWPA typically has losses.
Losses are especially harmful at the input of the TWPA,
before any amplification has taken place,
and it is imperative to both minimise the losses between the signal source and the TWPA \cite{AumentadoRev}
and to have a large ratio between the amplification per unit length and the damping per unit length, to suppress the internal losses.

Another property of TWPAs, which is rarely discussed, is the physical size.
While the size of a single TWPA is of small relevance in a large cryostat when there is only one TWPA,
the size becomes an issue when building large-scale quantum computers,
which would require many TWPAs.
A large-scale quantum computer may have qubits on the order of millions.
Even if each TWPA could multiplex up to 100 qubits, it follows that there is a need for TWPAs on the order of tens of thousands.
To fit all of these TWPAs inside one cryostat, there is hence a demand for TWPAs with a small physical footprint.

In this paper we demonstrate a TWPA implementing 3WM with the Superconducting Nonlinear Asymmetric Inductive eLement (SNAIL) \cite{Frattini2017}.
The TWPA uses a small cutoff frequency, in relation to the pump frequency, to inhibit up-conversion,
and a gap in the frequency spectrum to provide down-conversion phase matching \cite{MyTWPA3WM}.
We add impedance matching networks at the input and output ports to ensure impedance matching \cite{RenbergNilsson2024}.
By designing the TWPA this way, there is a high gain per unitcell,
which allows us to achieve a high overall gain of about 19\,dB for a chain of only 200 unitcells.
The high gain in a short length, in turn, allows us to
maximise the gain in relation to the losses per unit length,
ensuring that the losses of the TWPA have a small impact on the Signal-to-Noise Ratio (SNR).
Moreover, with the short length, the TWPA can have a straight design and there is no need for meandering.

\onecolumngrid

\begin{figure}[H]
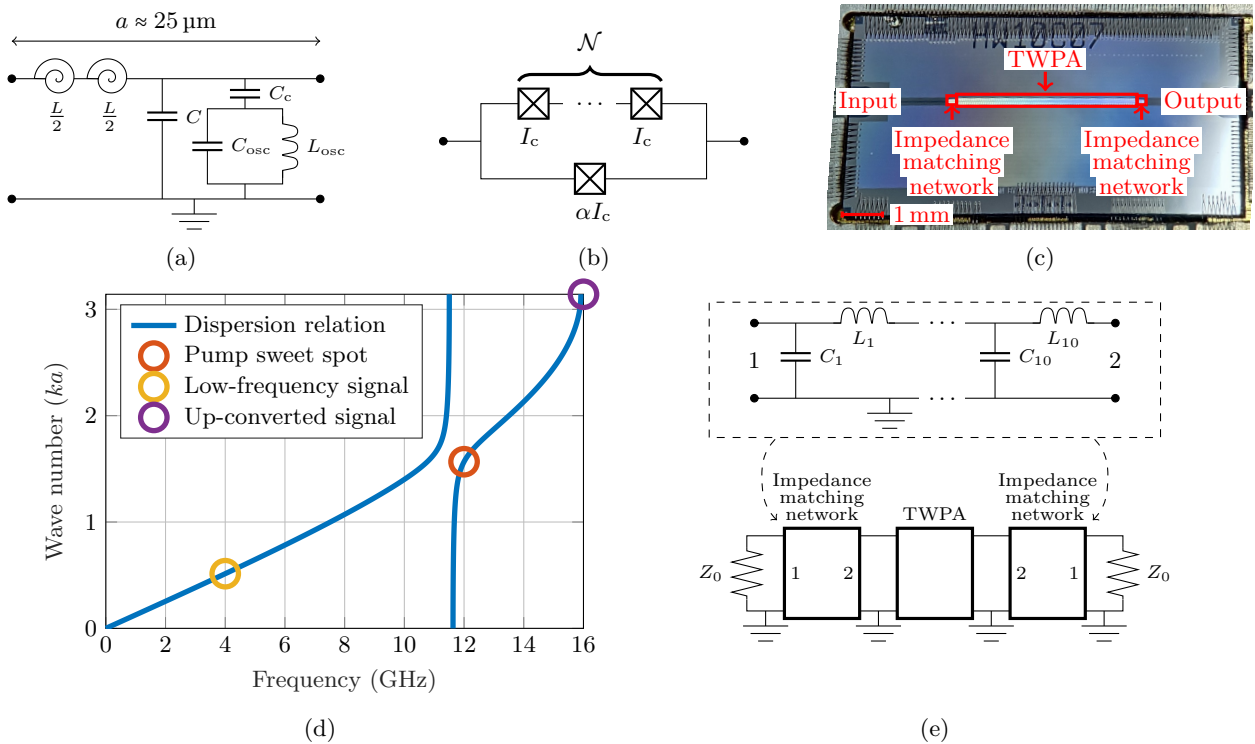

    \centering
    \begin{subfigure}{0.3\textwidth}
        \centering
        \tikzset{external/export next=false} 
\begin{tikzpicture}\draw[<->] (-0.3,1.5) -- (3.8,1.5) node[midway,anchor=south]{\(a\approx25\)\,\textmu{}m};

\foreach\y in {1,-0.6} {\draw (-0.3,\y) -- (3.8,\y);}
\SNAIL{0.3}1
\draw (0.3,0.5) node{\(\frac L2\)};
\SNAIL11
\draw (1,0.5) node{\(\frac L2\)};
\draw (1.7,1) -- +(0,-1.6);
\capacitor{1.7}{0.55}
\ground{2.125}{-0.6}
\draw (2.8,1) -- (2.8,-0.6);
\draw[fill=white] (2.3,-0.4) rectangle(3.3,0.6);
\capacitor[\ensuremath{C_\mathrm{c}}]{2.8}{0.85}
\capacitor[\ensuremath{C_\mathrm{osc}}]{2.3}{0.15}
\inductor[\ensuremath{L_\mathrm{osc}}]{3.3}{0.5}
\foreach\y in {-0.6,1} {
    \foreach\x in {-0.3,3.8} {\draw[fill] (\x,\y) circle(0.05);}
}\end{tikzpicture}
        \caption{}
        \label{fig:SNAIL_RPM_unitcell}
    \end{subfigure}
    \begin{subfigure}{0.3\textwidth}
        \centering
        \tikzset{external/export next=false} 
\begin{tikzpicture}
\draw[fill] (0,0.5) circle(0.05) -- (4,0.5) circle(0.05);
\draw[fill=white] (0.5,0) rectangle(3.5,1);
\fill[white] (1.6,0.9) rectangle(2.3,1.1);
\draw (1.95,1) node{\(\dots\)};

\JJ11
\JJ{2.5}1
\JJ{1.75}0

\draw[decorate,decoration={brace,amplitude=5pt},very thick] (1,1.3) -- (2.9,1.3);
\draw (1.95,1.6) node[anchor=south]{\(\mathcal N\)};

\draw (2,-0.2) node[anchor=north]{\(\alpha I\crit\)};
\foreach\x in {1,2.5} {
    \draw (\x+0.15,0.8) node[anchor=north]{\(I\crit\)};
}\end{tikzpicture}
        \caption{}
        \label{fig:SNAIL}
    \end{subfigure}
    \begin{subfigure}{0.35\textwidth}
        \centering
        \begin{tikzpicture}[scale=0.4]
            \input{tikzpics/Device2.tikz}
        \end{tikzpicture}
        \caption{}
        \label{fig:Device}
    \end{subfigure}
    \begin{subfigure}{0.45\textwidth}
        \centering
        \tikzset{external/export next=false} 
\begin{tikzpicture}\input{tikzpics/DispersionRelation.tikz}\end{tikzpicture}
        \caption{}
        \label{fig:Dispersion}
    \end{subfigure}
    \begin{subfigure}{0.45\textwidth}
        \centering
        \tikzset{external/export next=false} 
\begin{tikzpicture}\draw[white] (-3.25,-3) rectangle(3.25,2.5);

\draw (0,0.5) node[anchor=south]{
\begin{tikzpicture}

\foreach\y in {0,1} {
    \draw[fill] (0.2,\y) circle(0.05) -- (5,\y) circle(0.05);
    \fill[white] (2.4,\y-0.1) rectangle(3,\y+0.1);
    \draw (2.7,\y) node[anchor=center]{\(\dots\)};
}
\draw (0.2,0.5) node[anchor=center]{1};
\draw (5,0.5) node[anchor=center]{2};
\ground20

\draw (0.75,0) -- +(0,1);
\capacitor[C_1]{0.75}{0.6}
\hinductor[L_1]{1.25}{1}

\draw (3.4,0) -- +(0,1);
\capacitor[C_{10}]{3.4}{0.6}
\hinductor[L_{10}]{3.9}{1}

\end{tikzpicture}
};

\draw (0,0.25) node[anchor=north]{
\begin{tikzpicture}

\draw (0,0) rectangle(5,1);
\foreach\x in {0.25,1.75,3.25,4.75} {
    \ground\x0
}

\foreach\x in {0,5} {
    \vresistor\x{0.85}
}
\draw (-0.2,0.5) node[anchor=east]{\scriptsize\(Z_0\)};
\draw (5.2,0.5) node[anchor=west]{\scriptsize\(Z_0\)};

\foreach\x in {1,2.5,4} {
    \draw[very thick,fill=white] (\x-0.5,-0.1) rectangle(\x+0.5,1.1);
}

\foreach\x in {1,4} {
    \draw (\x,1.5) node[anchor=south]{\scriptsize Impedance} node[anchor=center]{\scriptsize matching} node[anchor=north]{\scriptsize network};
}
\draw (2.5,1.5) node{\scriptsize TWPA};

\foreach\x in {0.65,4.35} {
    \draw (\x,0.5) node[anchor=center]{\scriptsize1};
}
\foreach\x in {1.35,3.65} {
    \draw (\x,0.5) node[anchor=center]{\scriptsize2};
}

\end{tikzpicture}
};

\draw[dashed] (-3,0.5) rectangle(3,2.3);
\draw[dashed,<-] (-2.1,-0.6) to[out=135,in=-90] (-2.3,0) to[out=90,in=-135] (-2.1,0.5);
\draw[dashed,<-] (2.1,-0.6) to[out=45,in=-90] (2.3,0) to[out=90,in=-45] (2.1,0.5);\end{tikzpicture}
        \caption{}
        \label{fig:ImpMat}
    \end{subfigure}
    \caption{
        \textbf{Design elements, device and dispersion.}
        \Csref{fig:SNAIL_RPM_unitcell} The experimentally tested SNAIL-RPM-TWPA unitcell 
        with two SNAILs and a total series inductance \(L\),
        shunt capacitance \(C\), RPM coupling capacitance \(C\crit\), RPM oscillator capacitance \(C_\text{osc}\)
        and inductance \(L_\text{osc}\).
        \Csref{fig:SNAIL} A generic SNAIL,
        with \(\mathcal N\) Josephson junctions with the critical current \(I\crit\) in the upper arm
        and one junction with the critical current \(\alpha I\crit\) in the lower arm.
        \Csref{fig:Device}
        The measured TWPA with its impedance matching networks
        on a silicon substrate with the dimensions 5\,mm\(\times\)10\,mm.
        \Csref{fig:Dispersion}
        The dispersion relation:
        The wave number \(k\) times the unit cell length \(a\)
        as a function of frequency with a small signal frequency \(\omega\signal\) (yellow),
        the pump frequency sweet spot \(\omega\pump\) (orange)
        and the up-converted signal frequency \(\omega\pump+\omega\signal\) (purple) shown.
        \Csref{fig:ImpMat}
        The impedance matching networks placed at the input and the output of the TWPA.
        Each impedance matching network consists of 10 capacitors and 10 inductors.
        The output impedance matching network is the same as the one at the input,
        but with the components in reverse order.
    }
    \label{fig:Basics}
\end{figure}

\twocolumngrid

\section{TWPA design}
\label{sec:Basics}

The general structure of a TWPA is a chain of identical unitcells,
where each unitcell has a nonlinear series inductance and a linear shunt admittance,
although there are alternative designs \cite{Tien1958,Kow2024}.
In the simplest designs, the shunt admittance is a capacitor.
Then there are only two degrees of freedom, the series inductance \(L\) and the shunt capacitance \(C\).
More suitable quantities for the TWPA are the cutoff frequency \(\omega\crit = 2/\sqrt{LC}\) and the small-frequency impedance \(Z = \sqrt{L/C}\).
For small frequencies, \(\omega\ll\omega\crit\),
the TWPA impedance is equal to the small-frequency impedance,
while for large frequencies, \(\omega\sim\omega\crit\),
the TWPA impedance becomes complex-valued and frequency-dependent \cite{RenbergNilsson2024}.

To get a well working TWPA, three criteria need to be fulfilled:
impedance matching, down-conversion phase matching and up-conversion phase mismatching.
Impedance matching, \ie\ ensuring that the TWPA impedance is equal to the impedance \(Z_0\) of the environment for the relevant frequencies,
is necessary to prevent amplifier instability and to minimise gain ripples.
Down-conversion phase matching is required to ensure efficient energy transfer from the pump to the signal and the idler.
Up-conversion phase mismatching is necessary to suppress parasitic energy transfer from the pump, the signal and the idler to modes at higher frequencies.

Most 3WM TWPA proposals work on solving the issues described above
by choosing small operation frequencies compared to the spectral cutoff \cite{Pappas2016,GaoPRXQ2021,Zorin2016,AnitasArticle,Gaydamachenko2022}.
In this region the dispersion relation is almost linear,
which ensures good down-conversion phase matching.
Furthermore, the TWPA impedance is only weakly frequency-dependent,
which ensures impedance matching by setting the small-frequency TWPA impedance equal to the environment impedance, \(Z=Z_0\) \cite{RenbergNilsson2024}.
The remaining issue to enable exponential gain is up-conversion phase mismatching,
which has previously been done by creating a gap in the frequency spectrum at twice the pump frequency \cite{AnitasArticle,Gaydamachenko2022}.
However, in the small frequency regime,
the gain per unitcell is small \cite{MyTWPA3WM}.
Small gain per unitcell has two drawbacks:
i) the TWPA needs many unitcells in order to provide a large gain,
and ii) the losses of the TWPA may have a larger impact on the added noise.

To meet the three criteria formulated above, we instead take the approach proposed in Refs.~\cite{MyTWPA3WM,RenbergNilsson2024}:
We design the cutoff frequency \(\omega\crit\) to be close to the pump frequency \(\omega\pump\),
such that \(\omega\pump \sim 3\omega\crit/4\), see \Cref{fig:Dispersion},
in order to suppress the up-conversion processes.
At the same time, the efficiency of down-conversion is reduced for feasible pump powers.
To overcome this difficulty, we introduce a gap in the frequency spectrum below the pump frequency
by adding weakly coupled resonators in each unitcell, see \Cref{fig:SNAIL_RPM_unitcell}.
By settings \(C = C_\text{osc} = 10C\crit\),
we ensure that resonators affect the dispersion relation
mostly in a small spectral gap around the resonance frequency \(\omega\res\),
where \(\omega\res^{-2} = L_\text{osc}(C\crit + C_\text{osc})\).
The gap creates a sweet spot for the pump where its wave number is decreased,
which ensures phase matching in a band around half of the pump frequency \cite{MyTWPA3WM}, see \Cref{fig:Dispersion}.
    
By adding impedance matching networks to the input and the output ports, see \Cref{fig:ImpMat},
we also ensure impedance matching in a large frequency range of the TWPA \cite{RenbergNilsson2024}.
The impedance matching networks are constructed from
10 inductors and 10 capacitors
following the pattern of a Chebyshev filter \cite{MatthaeiYoungJones} of order 1000 and a ripple level of \(10^{-5}\)\,dB \cite{HRNthesis}.

We use the SNAIL as the nonlinear inductive element, 
which consists of a superconducting loop with several Josephson junctions, see \Cref{fig:SNAIL}.
The SNAIL characteristics are determined by
the critical current \(I\crit\),
the number of Josephson junctions \(\mathcal N\)
and the critical current ratio \(\alpha\leq1/\mathcal N\).
For our device, we have used SNAILs with
\(\mathcal N = 2\), \(\alpha = 0.25\) and \(I\crit = 1.8\)\,\textmu{}A.

Using a SNAIL is advantageous since there exists a magnetic flux-bias \(\phi_3 = \Phi_\mathrm{b}/\Phi_0 \in [0.25,0.5]\)
where 4WM Kerr effect is completely eliminated and we get pure three-wave mixing.
Furthermore, a SNAIL is based on Josephson junctions, which lets us have a large amount of inductance per unit area.
The inductance at the Kerr-free point is given by \cite{HRNthesis}
\begin{equation}
    L(\phi_3) = \frac{\mathcal N L\J[0]}{\displaystyle1 - \mathcal N^{-2}} \cdot \sqrt{\frac{1-\mathcal N^{-6}}{1-\alpha^2}}
    \approx \frac{\mathcal N L\J[0]}{\displaystyle1 - \mathcal N^{-2}},
\label{eq:SNAILinductance_atPhi3}
\end{equation}
where \(L\J[0] = \varphi_0 / I\crit\) is the inductance of a single junction at zero bias.
The 3WM coefficient \(\chi_3\) at the Kerr-free point, which describes the 3WM nonlinearity, is given by~\cite{HRNthesis}
\begin{equation}
    \chi_3(\phi_3) = \alpha\mathcal N \sqrt{\frac{1 - \alpha^{-2}\mathcal N^{-6}}{1 - \alpha^2}}.
\label{eq:SNAILc3_atPhi3}
\end{equation}

We aim for a pump sweet spot at 12\,GHz, which implies that the cutoff frequency should be around 16\,GHz.
This in turn implies that
the series inductance needs to be \(L \sim 1\)\,nH at the bias point,
and the total shunt capacitance \(C+C\crit \sim 400\)\,fF.

We use two SNAILs in series per unitcell.
Each of the two SNAILs have 2 larger junctions in one arm and one smaller junction in the other arm.
The larger junctions in the SNAIL are designed to have a critical current of \(\sim1.8\)\,\textmu{}A
and the smaller junction is designed to have a critical current of \(\sim450\)\,nA.

Using two SNAILS allows us to use junctions with twice larger critical current,
thus increasing the pump saturation power by 6\,dB.
The higher the saturation power, the more qubits may be multiplexed,
but this is at the cost of a larger physical size.
If we had used a single SNAIL per unitcell, the total length would almost be a factor 2 smaller.
However, since we already have such a small footprint, we decided to prioritize the higher saturation power.

\section{Fabrication}
\label{sec:Fabrication}

The device is fabricated on a silicon substrate with an aluminium wiring layer.
The junctions inside the SNAILs are also made of aluminium, deposited with shadow evaporation.

The capacitors are made with a parallel-plate geometry and a 40\,nm thin film of aluminium nitride (AlN) as the dielectric,
grown with atomic layer deposition (ALD).
Aluminium nitride has a dielectric constant of \(\varepsilon_\mathrm{r} \sim 7\) \cite{Khor2003}.
To get the correct capacitance, the capacitor \(C\) in each unitcell has an area of \(\sim260\)\,\textmu{}m\(^2\).

With the SNAILs and the parallel-plate capacitors, the physical footprint of the TWPA becomes very small.
The size of each impedance matching network is \(\sim0.02\)\,mm\(^2\),
and the size of a TWPA with 200 unitcells is \(\sim0.71\)\,mm\(^2\).
The total size of the TWPA is thus \(\sim0.75\)\,mm\(^2\), see \Cref{fig:Device}.
With the low number of unitcells, meandering of the TWPA is not needed,
which can further increase the density of TWPAs on a chip with several TWPAs.

However, if we place several TWPAs on the same chip, there needs to be some ground plane spacing between them.
If we add a ground plane of half of the width of RPM feature to each side of the TWPA,
several TWPAs on the same chip would be separated by a full RPM feature distance.
Then the TWPA footprint grows with \(\sim0.35\)\,mm\(^2\),
and the total footprint of the device thus becomes \(\sim1.1\)\,mm\(^2\).

\section{Gain Characterisation}
\label{sec:Characterisation}

The TWPA is characterised in a dilution refrigerator at 10\,mK,
see \Cref{fig:Setup} in \Cref{app:Setup} for details.
It is magnetically flux-biased by applying a current to a coil next to it.
It is placed between two cryogenic switches in parallel with a regular cable to use as a reference, the `through'.
After the TWPA, or the through, there is a state-of-the-art high-electron-mobility transistor (HEMT) amplifier.
When using the through, the dominant source of noise is the noise added by the HEMT amplifier.

\onecolumngrid

\begin{figure}[H]
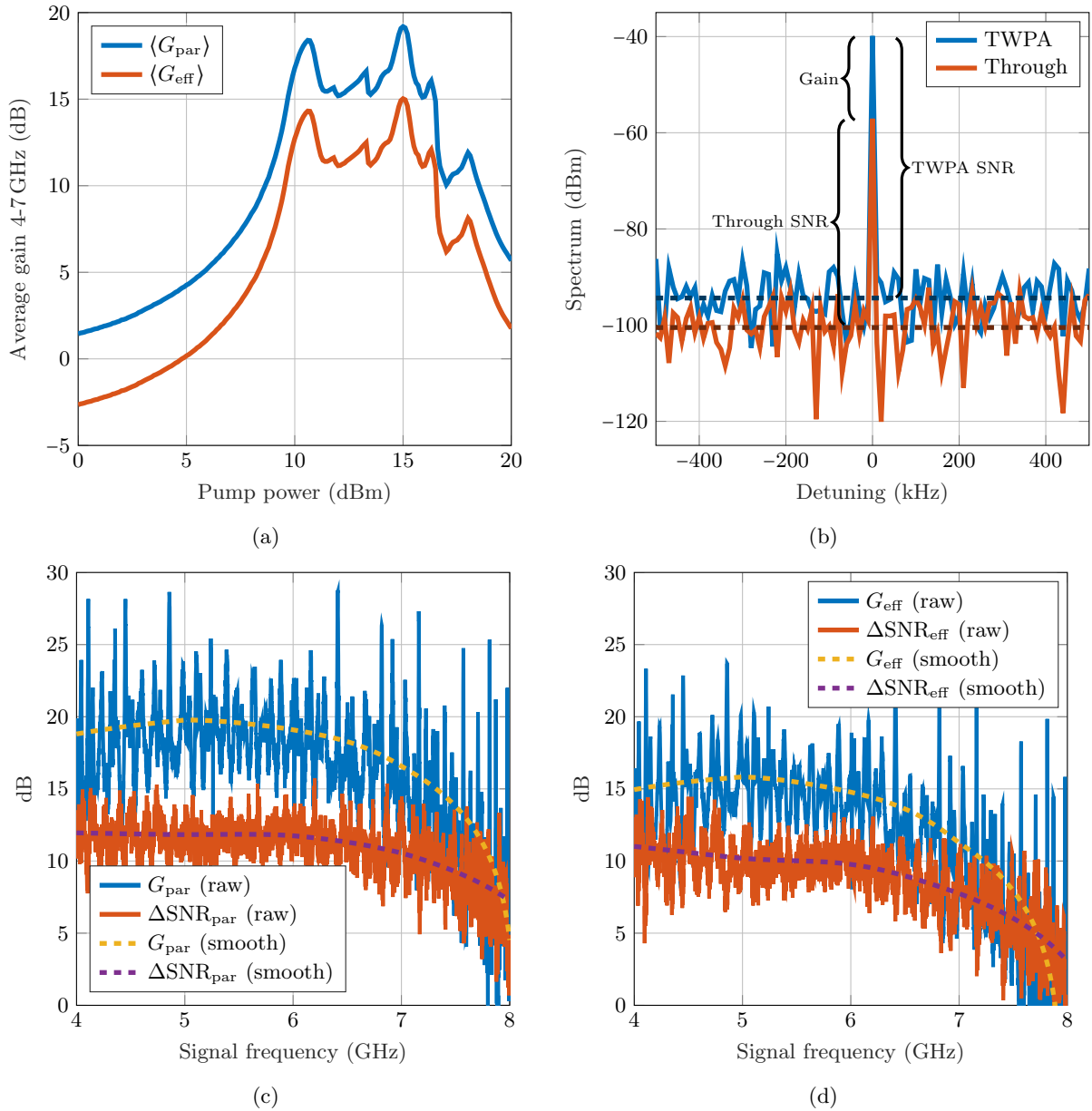

    \centering
    \begin{subfigure}{0.45\textwidth}
        \centering
        \tikzset{external/export next=false} 
\begin{tikzpicture}\input{tikzpics/AverageGain.tikz}\end{tikzpicture}
        \caption{}
        \label{fig:AvgGain}
    \end{subfigure}
    \begin{subfigure}{0.45\textwidth}
        \centering
        \tikzset{external/export next=false} 
\begin{tikzpicture}
%
%
\definecolor{mycolor1}{rgb}{0.00000,0.44700,0.74100}%
\definecolor{mycolor2}{rgb}{0.85000,0.32500,0.09800}%
\definecolor{mycolor3}{rgb}{0.00000,0.22350,0.37050}%
\definecolor{mycolor4}{rgb}{0.42500,0.16250,0.04900}%
%

\begin{axis}[%
width=2.5in,
height=2.5in,
at={(0.781in,0.542in)},
scale only axis,
xmin=-500,
xmax=500,
xlabel style={font=\color{white!15!black}},
xlabel={Detuning (kHz)},
ymin=-125,
ymax=-35,
ylabel style={font=\color{white!15!black}},
ylabel={Spectrum (dBm)},
axis background/.style={fill=white},
xmajorgrids,
ymajorgrids,
legend style={legend cell align=left, align=left, draw=white!15!black}
]
\addplot [color=mycolor1, line width=2.0pt]
  table[row sep=crcr]{%
-500	-86.234001159668\\
-490	-100.551002502441\\
-480	-100.453002929688\\
-470	-88.1829986572266\\
-460	-92.1029968261719\\
-450	-94.5199966430664\\
-440	-94.0989990234375\\
-430	-91.9710006713867\\
-420	-94.0609970092773\\
-410	-92.2210006713867\\
-400	-98.7389984130859\\
-390	-95.7470016479492\\
-380	-98.1060028076172\\
-370	-93.8489990234375\\
-360	-96.9670028686523\\
-350	-97.2080001831055\\
-340	-92.9140014648438\\
-330	-90.8529968261719\\
-320	-90.5189971923828\\
-310	-94.5130004882812\\
-300	-86.7910003662109\\
-290	-91.8560028076172\\
-280	-104.717002868652\\
-270	-96.7040023803711\\
-260	-96.0400009155273\\
-250	-89.1060028076172\\
-240	-89.4169998168945\\
-230	-104.443000793457\\
-220	-85.1890029907227\\
-210	-90.7170028686523\\
-200	-88.0419998168945\\
-190	-91.963996887207\\
-180	-96.322998046875\\
-170	-89.4530029296875\\
-160	-93.8990020751953\\
-150	-94.4909973144531\\
-140	-96.9229965209961\\
-130	-97.1699981689453\\
-120	-96.870002746582\\
-110	-97.5810012817383\\
-100	-88.9130020141602\\
-90	-88.0130004882812\\
-80	-89.8639984130859\\
-70	-97.3919982910156\\
-60	-103.78099822998\\
-50	-99.1529998779297\\
-40	-91.1630020141602\\
-30	-96.9169998168945\\
-20	-100.305000305176\\
-10	-95.4039993286133\\
0	-39.875\\
10	-89.9069976806641\\
20	-91.2070007324219\\
30	-95.1389999389648\\
40	-94.6719970703125\\
50	-89.0790023803711\\
60	-90.4100036621094\\
70	-99.8190002441406\\
80	-103.502998352051\\
90	-103.180000305176\\
100	-88.3779983520508\\
110	-95.4059982299805\\
120	-90.2839965820312\\
130	-90.661003112793\\
140	-102.188003540039\\
150	-87.3960037231445\\
160	-97.5\\
170	-90.5859985351562\\
180	-94.2809982299805\\
190	-99.8249969482422\\
200	-98.0240020751953\\
210	-89.2910003662109\\
220	-97.2289962768555\\
230	-91.3239974975586\\
240	-100.125\\
250	-101.539001464844\\
260	-95.7679977416992\\
270	-98.4820022583008\\
280	-96.484001159668\\
290	-92.7269973754883\\
300	-89.8410034179688\\
310	-92.5090026855469\\
320	-95.9919967651367\\
330	-99.9919967651367\\
340	-97.0199966430664\\
350	-92.1829986572266\\
360	-91.1230010986328\\
370	-96.984001159668\\
380	-90.8889999389648\\
390	-91.9749984741211\\
400	-97.568000793457\\
410	-92.1890029907227\\
420	-87.2279968261719\\
430	-88.322998046875\\
440	-102.305000305176\\
450	-90.7669982910156\\
460	-92.786003112793\\
470	-96.963996887207\\
480	-98.8249969482422\\
490	-95.8130035400391\\
500	-88.2249984741211\\
};
\addlegendentry{TWPA}

\addplot [color=mycolor2, line width=2.0pt]
  table[row sep=crcr]{%
-500	-101.708999633789\\
-490	-102.608001708984\\
-480	-98.5640029907227\\
-470	-107.871002197266\\
-460	-97.4820022583008\\
-450	-98.3990020751953\\
-440	-103.259002685547\\
-430	-98.629997253418\\
-420	-96.3079986572266\\
-410	-101.579002380371\\
-400	-98.3649978637695\\
-390	-102.694000244141\\
-380	-99.6240005493164\\
-370	-99.4489974975586\\
-360	-103.609001159668\\
-350	-105.023002624512\\
-340	-108.453002929688\\
-330	-104.818000793457\\
-320	-96.5090026855469\\
-310	-98.4250030517578\\
-300	-94.7880020141602\\
-290	-101.009002685547\\
-280	-101.838996887207\\
-270	-100.972999572754\\
-260	-104.658996582031\\
-250	-98.2239990234375\\
-240	-96.2460021972656\\
-230	-94\\
-220	-106.63200378418\\
-210	-94.2320022583008\\
-200	-97.0830001831055\\
-190	-93.0410003662109\\
-180	-101.269996643066\\
-170	-103.563003540039\\
-160	-98.9660034179688\\
-150	-100.829002380371\\
-140	-96.7559967041016\\
-130	-119.564002990723\\
-120	-96.3010025024414\\
-110	-101.329002380371\\
-100	-100.160003662109\\
-90	-104.301002502441\\
-80	-93.5960006713867\\
-70	-109.81600189209\\
-60	-103.945999145508\\
-50	-102.200996398926\\
-40	-95.7460021972656\\
-30	-101.986000061035\\
-20	-97.036003112793\\
-10	-99.0059967041016\\
0	-57.0670013427734\\
10	-103.722999572754\\
20	-120.08699798584\\
30	-98.0910034179688\\
40	-98.1100006103516\\
50	-97.8639984130859\\
60	-109.700996398926\\
70	-105.166999816895\\
80	-96.7040023803711\\
90	-96.3239974975586\\
100	-94.4879989624023\\
110	-96.3629989624023\\
120	-99.8820037841797\\
130	-92.1999969482422\\
140	-104.245002746582\\
150	-101.564002990723\\
160	-109.684997558594\\
170	-104.949996948242\\
180	-100.453002929688\\
190	-95.1039962768555\\
200	-95.6869964599609\\
210	-113.03099822998\\
220	-99.1439971923828\\
230	-92.9820022583008\\
240	-99.7649993896484\\
250	-96.2379989624023\\
260	-100.140998840332\\
270	-95.4749984741211\\
280	-95.3769989013672\\
290	-93.4349975585938\\
300	-94.2279968261719\\
310	-106.345001220703\\
320	-95.2900009155273\\
330	-103.853996276855\\
340	-92.9609985351562\\
350	-98.1399993896484\\
360	-98.6969985961914\\
370	-98.4039993286133\\
380	-101.831001281738\\
390	-97.1100006103516\\
400	-100.805000305176\\
410	-102.33699798584\\
420	-100.380996704102\\
430	-110.276000976562\\
440	-118.286003112793\\
450	-103.278999328613\\
460	-95.1220016479492\\
470	-102.241996765137\\
480	-103.383003234863\\
490	-94.8870010375977\\
500	-93.6880035400391\\
};
\addlegendentry{Through}

\addplot [color=mycolor3, dashed, forget plot, line width=2.0pt]
  table[row sep=crcr]{%
-500	-94.3632699584961\\
500	-94.3632699584961\\
};
\addplot [color=mycolor4, dashed, forget plot, line width=2.0pt]
  table[row sep=crcr]{%
-500	-100.500020217896\\
500	-100.500020217896\\
};
\end{axis}

\draw[decorate,decoration={brace,amplitude=5pt},very thick] (4.75,3.15) -- +(0,3) node[midway,anchor=east,left=0.1]{\scriptsize Through SNR};

\draw[decorate,decoration={brace,amplitude=5pt},very thick] (4.9,3.15+3) -- +(0,1.22) node[midway,anchor=east,left=0.1]{\scriptsize Gain};

\draw[decorate,decoration={brace,amplitude=5pt},very thick] (5.5,7.35) -- +(0,-3.8) node[midway,anchor=west,right=0.1]{\scriptsize TWPA SNR};\end{tikzpicture}
        \caption{}
        \label{fig:SNR}
    \end{subfigure}
    \begin{subfigure}{0.45\textwidth}
        \centering
        \tikzset{external/export next=false} 
\begin{tikzpicture}\input{tikzpics/ParGainAndDeltaSNR.tikz}\end{tikzpicture}
        \caption{}
        \label{fig:ParGain}
    \end{subfigure}
    \begin{subfigure}{0.45\textwidth}
        \centering
        \tikzset{external/export next=false} 
\begin{tikzpicture}\input{tikzpics/EffGainAndDeltaSNR.tikz}\end{tikzpicture}
        \caption{}
        \label{fig:EffGain}
    \end{subfigure}
    \caption{
        \textbf{Measurements.}
        Measurements with a pump at \(f\pump=11.27\)\,GHz and a flux bias \(\phi \approx \phi_3\).
        \Csref{fig:AvgGain}
        The arithmetic averages of the gain in the interval 4-7\,GHz,
        both with and without the TWPA losses,
        for different room temperature pump powers.
        \Csref{fig:SNR}
        One of the SNR measurements.
        The noise floors are determined as the geometric averages of all data points at nonzero detuning.
        The SNR is determined as the difference in dB between the signal at zero detuning and the noise floor.
        \Csref{fig:ParGain}
        The parametric gain (\(G_\text{par}\)) and the parametric Signal-to-Noise Ratio improvement (\(\Delta\text{SNR}_\text{par}\)),
        with smooth fits.
        \Csref{fig:EffGain}
        The effective gain (\(G_\text{eff}\)) and the effective Signal-to-Noise Ratio improvement (\(\Delta\text{SNR}_\text{eff}\)),
        with smooth fits.
        The smooth curves are generated using the `smooth' function in MATLAB,
        with the input arguments `lowess' and 0.5.
    }
    \label{fig:Measurements}
\end{figure}

\twocolumngrid 

To measure the gain, we measure the transmission with a vector network analyzer (VNA)
while we sweep the flux, the pump frequency and the pump amplitude.
The parametric gain \(G_\text{par}\) is defined as the transmission of the TWPA when the pump is on 
in relation to the transmission of the TWPA when the pump is off.
The effective gain \(G_\text{eff}\) is defined as the transmission of the TWPA when the pump is on
in relation to when the signal is routed via the through.
The latter is a more accurate measure, since it takes the losses of the TWPA into consideration,
assuming the loss of the short superconducting `through' wire is negligible.

The arithmetic averages of the gain \(\langle G \rangle\) in the interval 4-7\,GHz are presented in \Cref{fig:AvgGain}.
As we can see, the maximum average gain is found at the room temperature pump power \(P\pump=15\)\,dBm.
The plot is expressed in terms of room temperature pump power,
since we do not know the exact amount of loss of the line,
but we estimate that the full attenuation of the pump
from the signal generator to the TWPA chip,
including room temperature attenuators and the directional couplers,
is approximately 103\dB.

The measured gain for different frequencies and the pump power \(P\pump=15\dBm\) is presented in \Cref{fig:ParGain,fig:EffGain}.
The average gain excluding losses, \(G_\text{par}\), in the interval 4 to 7\,GHz is 19\,dB.
The average is determined as the arithmetic average of the power gain.
When including losses, which are approximately 4\,dB in the interval 4-7\,GHz,
the average effective gain \(G_\text{eff}\) drops to 15\,dB.

Up to the pump power \(P\pump=10\dBm\), the gain grows exponentially, as expected from theory, see \Cref{fig:AvgGain}.
At larger pump powers the gain exhibits an irregular behaviour, instead of exponential growth,
and after reaching a second peak at \(P\pump=15\dBm\), it drops down. 
We interpret the latter peak as the switching point where some of the SNAILs lose stability and switch to the resistive state.

A quantitative comparison between the data and theory in Refs.~\cite{MyTWPA3WM,RenbergNilsson2024} is difficult due to
the lack of knowledge of the exact values of the on-chip pump power and switching current,
and the TWPA gain is very sensitive to pump power.
Furthermore, the calculation of gain was done for a lossless TWPA, although for a device with a similar design and length.
Therefore we make comparison with the \(G_\text{par}\) data with caution.

The average gain at \(P\pump=10\dBm\) is \(\langle G_\text{par} \rangle = 17\dB\),
which corresponds to the gain coefficient \(g\approx0.01\) in accordance with the formula
\begin{equation}
    G_\text{dB} = 20gN\log_{10}(\mathrm{e})
\label{eq:GainCoeff}
\end{equation}
where \(N\) is the number of unitcells
and \(G_\text{dB}\) is the gain expressed in dB \cite{MyTWPA3WM}.
According to the calculation, this gain is achieved for the pump current \(I\pump \approx 0.084 I\crit\)
for the SNAIL 3WM coefficient at the working point, \(\chi_3\approx0.5\).
The corresponding on-chip pump power is \(P\pump \approx - 91\dBm\),
which implies a total pumpline attenuation of 101\dB,
which is very close to the estimated amount of attenuation.
For this pumping strength, theory predicts the gain bandwidth
\(\approx 0.4\omega\pump = 4.5\)\,GHz.
This seems to be consistent with the data,
although we cannot measure the full gain bandwidth
because of the bandwidths of both the HEMT and the isolators.

Adopting these calculations, we get that the switching current is \(I_\text{sw} \approx 0.17 I\crit\), 
which implies that the switching would occur rather prematurely.
The reason for such a premature switching, as well as the gain irregularity within the interval \(P\pump=10\) to \(15\dBm\), remains unclear.

\section{Signal-to-Noise Ratio Characterisation}
\label{sec:SNR}

While the gain is an important metric, more important metrics in the context of low noise amplifiers
are the noise temperature, the quantum efficiency or the number of added noise photons.
However, these can be hard to calibrate in the full frequency range of the TWPA.
A metric that we can determine in the full frequency range of the TWPA is the Signal-to-Noise Ratio improvement (\(\Delta\)SNR).
While the \(\Delta\)SNR is setup-dependent, our output line is directly connected to the HEMT amplifier.
The SNR of our reference, the through, should hence be a good estimate of
the best achievable SNR without a quantum-limited amplifier.

To determine the \(\Delta\)SNR, we insert a probe tone
and measure the output within 1\,MHz band around the probe, see \Cref{fig:SNR}.
We define the signal as the output at zero detuning,
the noise floor as the average value of the points at non-zero detuning
and the SNR as the difference between the signal and the noise floor.
The parametric \(\Delta\)SNR is defined as the SNR of the TWPA when the pump is on
in relation to the SNR of the TWPA when the pump is off.
The effective \(\Delta\)SNR is defined as the SNR of the TWPA when the pump is on
in relation to the SNR measured by the HEMT amplifier via the through.

The measured \(\Delta\)SNR is presented in \Cref{fig:ParGain,fig:EffGain}.
While the maximum average gain was found at \(P\pump=15\dBm\),
the maximum average \(\Delta\)SNR was found at \(P\pump=10\dBm\).
Therefore the gain plots in \Cref{fig:ParGain,fig:EffGain} use \(P\pump=15\dBm\)
but the \(\Delta\)SNR plots use \(P\pump=10\dBm\).
The average parametric \(\Delta\)SNR in the interval 4-7\,GHz for \(P\pump=10\dBm\) is 12\dB,
and the average effective \(\Delta\)SNR is 10\dB.
It is worth noting that the highest \(\Delta\)SNR is found at \(P\pump=10\dBm\),
which is also the highest pump power where the gain still grows exponentially.

\section{Noise model}
\label{sec:Noise}

We see that when we include the losses of the device, the average gain drops 4\dB,
while the average \(\Delta\)SNR only drops 2\dB.
A rough model to understand this can be made by assuming that
the TWPA is completely lossless,
and the 4\dB\ TWPA loss is placed in two attenuators,
2\dB\ in front of the TWPA, and 2\dB\ after the TWPA.

When the input signal reaches the first attenuator, it is attenuated with 2\dB.
However, the input noise is equal to the vacuum fluctuations,
and thus cannot be attenuated further.
Then the signal and the noise reach the lossless amplifier,
and the amplifier amplifies and adds noise to the extent that the parametric \(\Delta\)SNR equals 12\dB.
Then the signal and the noise reach the second 2\dB\ attenuator.
However, this time the noise has been amplified well above the level of vacuum fluctuations,
and thus both the signal and the noise floor are attenuated.

\onecolumngrid


\begin{figure}[H]
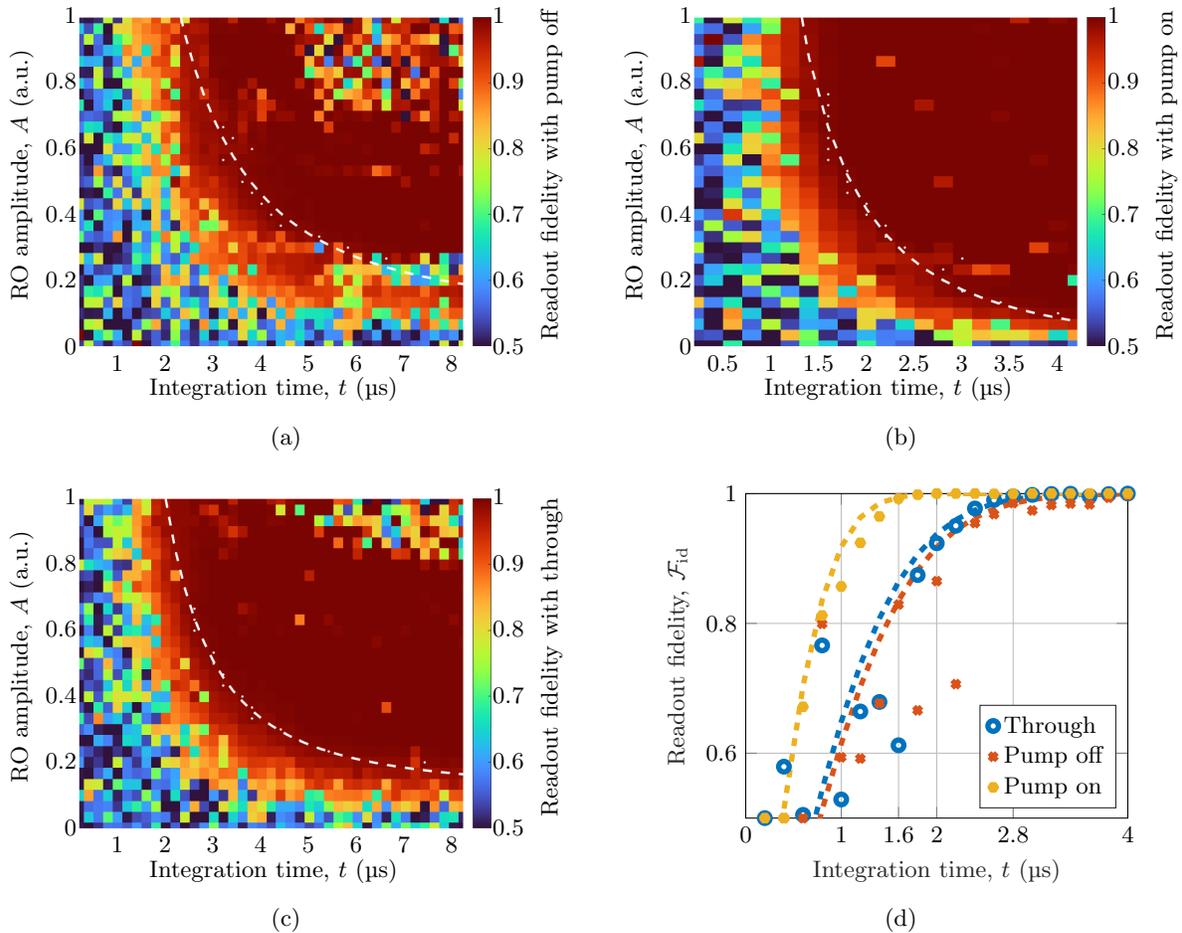

    \centering
    \begin{subfigure}{0.45\textwidth}
        \centering
        \tikzset{external/export next=false} 
\begin{tikzpicture}\input{tikzpics/qubitpics/PumpOff2d.tikz}\end{tikzpicture}
        \caption{}
        \label{fig:PumpOff2d}
    \end{subfigure}
    \begin{subfigure}{0.45\textwidth}
        \centering
        \tikzset{external/export next=false} 
\begin{tikzpicture}\input{tikzpics/qubitpics/PumpOn2d.tikz}\end{tikzpicture}
        \caption{}
        \label{fig:PumpOn2d}
    \end{subfigure}
    \begin{subfigure}{0.45\textwidth}
        \centering
        \tikzset{external/export next=false} 
\begin{tikzpicture}\input{tikzpics/qubitpics/Thru2d.tikz}\end{tikzpicture}
        \caption{}
        \label{fig:Thru2d}
    \end{subfigure}
    \begin{subfigure}{0.45\textwidth}
        \centering
        \tikzset{external/export next=false} 
\begin{tikzpicture}
%
%
\definecolor{mycolor1}{rgb}{0.00000,0.44700,0.74100}%
\definecolor{mycolor2}{rgb}{0.85000,0.32500,0.09800}%
\definecolor{mycolor3}{rgb}{0.92900,0.69400,0.12500}%
%

\begin{axis}[%
width=2in,
height=1.7in,
at={(0.838in,0.547in)},
scale only axis,
xmin=0,
xmax=4,
xtick={0,1,1.6,2,2.8,4},
xlabel style={font=\color{white!15!black}},
xlabel={Integration time, \(t\) (\textmu{}s)},
ymin=0.5,
ymax=1,
ylabel style={font=\color{white!15!black}},
ylabel={Readout fidelity, \(\mathcal F_\mathrm{id}\)},
axis background/.style={fill=white},
xmajorgrids,
ymajorgrids,
legend style={at={(0.97,0.03)}, anchor=south east, legend cell align=left, align=left, draw=white!15!black}
]
\addplot [color=mycolor1, line width=2.0pt, only marks, mark=o, mark options={solid, mycolor1}]
  table[row sep=crcr]{%
0.2	0.5\\
0.4	0.579268513052833\\
0.6	0.505007157219439\\
0.8	0.766086354300667\\
1	0.528702010578233\\
1.2	0.664266789476785\\
1.4	0.678954165339759\\
1.6	0.612242383728731\\
1.8	0.874481747190945\\
2	0.923466828393253\\
2.2	0.950383496044125\\
2.4	0.97703048290962\\
2.6	0.990294551725343\\
2.8	0.995609808294592\\
3	0.997955826173773\\
3.2	0.999356035050024\\
3.4	0.999766194198604\\
3.6	0.995841573245072\\
3.8	0.999230224712758\\
4	0.999978033459217\\
4.2	0.999985146497225\\
4.4	0.996592130448988\\
4.6	0.999997494205712\\
4.8	0.999877250424333\\
5	0.999999982709161\\
5.2	0.999949239855632\\
5.4	0.999967701886015\\
5.6	0.994251132713957\\
5.8	0.999961553840141\\
6	0.999890863489838\\
6.2	0.999959087792763\\
6.4	0.999924743728112\\
6.6	0.999922668844033\\
6.8	0.99998236741153\\
7	0.99566711264674\\
7.2	0.99999791255794\\
7.4	0.999999999739005\\
7.6	0.999999999997122\\
7.8	0.999999997966419\\
8	0.99999999990042\\
8.2	0.999851097207988\\
};
\addlegendentry{Through}

\addplot [color=mycolor2, line width=2.0pt, only marks, mark=x, mark options={solid, mycolor2}]
  table[row sep=crcr]{%
0.2	0.5\\
0.4	0.5\\
0.6	0.5\\
0.8	0.799255890435968\\
1	0.593623137642838\\
1.2	0.591812503310746\\
1.4	0.676748612674764\\
1.6	0.829043902035898\\
1.8	0.665962921670908\\
2	0.865075858708014\\
2.2	0.706404219988781\\
2.4	0.954533272120212\\
2.6	0.968222246146032\\
2.8	0.985206258209675\\
3	0.973964989284583\\
3.2	0.982426326948298\\
3.4	0.984523132554957\\
3.6	0.983361987687159\\
3.8	0.993607848406608\\
4	0.997030687123537\\
4.2	0.998220922138087\\
4.4	0.996999083576899\\
4.6	0.999461384952923\\
4.8	0.99957747109912\\
5	0.999952406383169\\
5.2	0.99996530246024\\
5.4	0.997925702338105\\
5.6	0.999993337945613\\
5.8	0.999979511388172\\
6	0.999997638148537\\
6.2	0.994305013577272\\
6.4	0.999999488686724\\
6.6	0.999999944159312\\
6.8	0.999999617358157\\
7	0.999999896043673\\
7.2	0.999999931543556\\
7.4	0.952528539918098\\
7.6	0.973437724691864\\
7.8	0.999983831063527\\
8	0.999995583392824\\
8.2	0.99999999901497\\
};
\addlegendentry{Pump off}

\addplot [color=mycolor3, line width=2.0pt, only marks, mark=asterisk, mark options={solid, mycolor3}]
  table[row sep=crcr]{%
0.2	0.5\\
0.4	0.5\\
0.6	0.671549162094052\\
0.8	0.812285512541287\\
1	0.856971010155109\\
1.2	0.92414424198166\\
1.4	0.964678131825299\\
1.6	0.992144261160967\\
1.8	0.998269710050923\\
2	0.999972298096905\\
2.2	0.999997685806514\\
2.4	0.99999986165939\\
2.6	0.999596712804316\\
2.8	0.999816710283591\\
3	0.999999999998875\\
3.2	0.999999999999997\\
3.4	0.999999999999999\\
3.6	1\\
3.8	1\\
4	1\\
4.2	1\\
};
\addlegendentry{Pump on}

\addplot [color=mycolor1, dashed, line width=2.0pt, forget plot]
  table[row sep=crcr]{%
0.2	0.147813917489409\\
0.4	0.290598300021814\\
0.6	0.423831194719555\\
0.8	0.543921399272679\\
1	0.648482237115769\\
1.2	0.736424138822226\\
1.4	0.807871789525183\\
1.6	0.863943696215705\\
1.8	0.906451427805747\\
2	0.937579751042848\\
2.2	0.95959931280174\\
2.4	0.974645508591122\\
2.6	0.984576892794366\\
2.8	0.990909139584341\\
3	0.994809195104578\\
3.2	0.997129518839106\\
3.4	0.998463013048495\\
3.6	0.999203297048321\\
3.8	0.999600278469223\\
4	0.999805917950514\\
4.2	0.999908816018428\\
};
\addplot [color=mycolor2, dashed, line width=2.0pt, forget plot]
  table[row sep=crcr]{%
0.2	0.137845302737504\\
0.4	0.27160697641904\\
0.6	0.397560736383935\\
0.8	0.512648802553952\\
1	0.614693197167897\\
1.2	0.702491820704546\\
1.4	0.775795527944297\\
1.6	0.835184229612374\\
1.8	0.881873959594888\\
2	0.917492696056575\\
2.2	0.943860577974394\\
2.4	0.962801954635966\\
2.6	0.976005402730824\\
2.8	0.984936455924512\\
3	0.990798612887428\\
3.2	0.994532417789007\\
3.4	0.996840148705099\\
3.6	0.998224218644004\\
3.8	0.999029727233077\\
4	0.999484633096077\\
4.2	0.999733927228501\\
};
\addplot [color=mycolor3, dashed, line width=2.0pt, forget plot]
  table[row sep=crcr]{%
0.2	0.27214311405255\\
0.4	0.51354331397312\\
0.6	0.703483740429556\\
0.8	0.836050661618179\\
1	0.918121454256132\\
1.2	0.963190126567506\\
1.4	0.985142922780785\\
1.6	0.994627750534117\\
1.8	0.998262617210199\\
2	0.999498169823785\\
2.2	0.999870682807508\\
2.4	0.999970297210636\\
2.6	0.999993923528268\\
2.8	0.999998893515527\\
3	0.999999820754294\\
3.2	0.999999974179356\\
3.4	0.999999996693761\\
3.6	0.999999999623809\\
3.8	0.999999999961975\\
4	0.999999999996587\\
4.2	0.999999999999728\\
};
\end{axis}
\end{tikzpicture}
        \caption{}
        \label{fig:Linecuts}
    \end{subfigure}
    \caption{
        \textbf{Qubit readout fidelity versus measurement time and readout amplitude.}
        Fits are shown with dashed lines.
        For the first three subfigures, the fit is of the function \(A\propto1/\sqrt{t}\),
        while for the last, the fit is of the function erf\((at)\).
        \Csref{fig:PumpOff2d}
        A 2D fidelity plot using the TWPA with the pump off.
        \Csref{fig:PumpOn2d}
        A 2D fidelity plot using the TWPA with the pump on.
        \Csref{fig:Thru2d}
        A 2D fidelity plot using the through.
        \Csref{fig:Linecuts}
        Measured readout fidelity versus integration time.
    }
    \label{fig:Qubit}
\end{figure}

\twocolumngrid

Therefore, the effective \(\Delta\)SNR is lowered with only half the losses,
which is what we saw experimentally.
While this model seems to explain the different effects of the losses on the gain versus the \(\Delta\)SNR,
it is a very simple model, and in reality there are also losses distributed within the TWPA.

If we assume that the TWPA consists of \(N\) segments,
and each segment has the gain \(\sqrt[N]G>1\) and the damping \(\sqrt[N]D<1\),
one can show that the total number of noise photons added by the TWPA is
\begin{equation}
    A_\text{TWPA} = \frac{GD - 1}{2GD} \frac{\pln{G/D}}{\pln{GD}},
\label{eq:A_TWPA}
\end{equation}
see \Cref{app:Noise} for the derivation of \Cref{eq:A_TWPA}.
However, this model assumes a homogeneous distribution of the losses.
This is likely true inside the TWPA, but there may also be losses in the connectors and the PCB.
To build an accurate model, we would need to determine the distribution of the losses,
inside the TWPA itself, the connectors and the PCB.


\section{Qubit measurements}
\label{sec:Qubits}

In order to characterize the improved readout performance from the TWPA,
we use the TWPA in the readout chain of a two-qubit device \cite{Kosen2022}.
The device consists of two fixed-frequency transmon qubits \cite{Koch2007}
with transition frequencies \(\omega_{q_{i}}/(2\pi)\) at 3.848 and 3.384\,GHz for \(i = 0\) and 1, respectively.
The qubits are coupled via a tunable coupler element, which is not used in this experiment,
thus we can consider the qubits uncoupled.

Each qubit is coupled to a dedicated readout resonator of frequency \(\omega_{r_{i}}/(2\pi) = 6.482\) and \(6.261\)\,GHz with a strength \(g/(2\pi) \approx 35\)\,MHz.
The transmitted signal is measured through the readout line.
We prepared the two qubits in \(\ket{g}\) and \(\ket{e}\) states respectively
and characterised their single-shot readout performance with multiplexed readout.
The measurement results of each qubit state are fitted with a Gaussian function
and the ideal readout fidelity \(\mathcal{F}_\text{id}\) was calculated as \cite{Magesan2015, Chen2023}
\begin{equation}
    \mathcal{F}_\text{id} =  \frac12 \left[ 1 + \mathrm{erf}\left(\sqrt{\frac{\text{SNR}^{2}}{8}}\right) \right],
\label{eq:fid}
\end{equation}
where the SNR is
\begin{equation}
    \text{SNR} = \frac{\left\lvert \left\langle S_0 \right\rangle - \left\langle S_{\overline{0}} \right\rangle \right\lvert}{\sigma_{S_0}}.
\label{eq:p1EffhPulse}
\end{equation}
with \(S_{i}\) being the set of measurement outcomes and \(\sigma^2_{S_{0}}\) being the variance of the data set.
\(1-\mathcal{F}_\text{id}\) corresponds to the error from the overlap of the Gaussian distribution of the two qubit states,
which is caused by the thermal noise of the readout chain and thus essentially quantifies the SNR of the system.

We set a fixed readout pulse amplitude and sweep the integration time of the readout data acquisition process.
The resulting readout fidelity for Qubit 1 with TWPA pump on, pump off and with a through 
is shown in \Cref{fig:PumpOff2d,fig:PumpOn2d,fig:Thru2d} as an example.
The total energy needed to achieve sufficient SNR is constant, which can be fitted with a functional form of \(E\propto\ t \cdot A^{2}\),
where \(A\) is the readout amplitude.
The linecut at a fixed amplitude is shown in \Cref{fig:Linecuts},
where the data can be approximately fitted with a Gaussian error function to extract the dependence between the readout fidelity and integration time \cite{Walter2017}.

To reach a readout fidelity of 99.9\,\%, it takes 2.8\,\textmu{}s of integration time with the through,
while we only need 1.6\,\textmu{}s with the TWPA with the pump on,
achieving almost a factor of two improvement in the readout speed.
With the pump off, the time increases to 3.0\,\textmu{}s.
It is worth noting that the performance is also more robust at the shorter integration time with the TWPA,
indicating that a shorter readout time is achievable.
With a device designed with a larger qubit-readout dispersive coupling strength \(\chi\), resonator decay strength \(\kappa\), and incorporating Purcell filters,
it is still possible to reduce the readout duration greatly \cite{Chen2023}.

\section{Discussion and Conclusions}
\label{sec:Conclusions}

In this paper we have demonstrated an efficient travelling-wave parametric amplifier with
a high signal-to-noise ratio improvement in a wide band and a small footprint.
The TWPA works with frequencies close to the cutoff frequency
and uses the cutoff frequency to inhibit up-conversion
and resonant phase matching to ensure down-conversion phase matching \cite{MyTWPA3WM,PatentOne}.
It uses impedance matching networks at the input and the output to ensure impedance matching \cite{RenbergNilsson2024}.
With a total physical footprint of 1.1\,mm\(^2\),
and only 200 unitcells,
we still achieve an average parametric gain of 19\,dB
and an average effective \(\Delta\)SNR of 10\,dB
over a 3\,GHz band,
and a clear speedup in qubit readout integration time.
Furthermore, the gain, and in turn the \(\Delta\)SNR, has a fairly flat shape over several GHz of bandwidth.

The device has several dB of insertion loss,
which makes a 4\,dB average difference on the gain.
Yet, these losses have a smaller impact on the Signal-to-Noise Ratio improvement,
which is only lowered with 2\,dB on average.
We attribute this small difference of the \(\Delta\)SNR to the high gain per unitcell
compared to the internal losses of the TWPA,
thus making the internal losses of the TWPA less significant.

In conclusion, we have demonstrated a device we believe will be advantageous when building large-scale quantum computers,
due to its small physical footprint and high Signal-to-Noise Ratio improvement.

\section{Acknowledgements}
The project was supported by the Knut and Alice Wallenberg foundation via the Wallenberg Centre for Quantum Technology.
The authors acknowledge the use of the Nanofabrication Laboratory (NFL) at Chalmers University of Technology, and the help of the NFL staff.


\appendix

\onecolumngrid
\clearpage
\twocolumngrid

\section{Experimental setup}
\label{app:Setup}

\begin{figure}[H]
    \centering
    \tikzset{external/export next=false} 
\begin{tikzpicture}\draw (1.2,4) -- (1.2,4.5) node[anchor=south]{\scriptsize Signal in 1};
\simptermuddircoup{0}{3.5}

\draw (-1,3.5) -- (0,3.5);
\draw (-1,4) node{\scriptsize Pump in};
\generator{-1}{3.5}

\draw[dashed] (2.5,0.5) rectangle(13.5,6.4);
\draw (2.5,0.5) node[anchor=north west]{\scriptsize Dilution refrigerator};

\draw (1.5,3.5) -- (9.7,3.5) -- +(0,-1.5);

\draw (10.5,1.2) -- (12,1.2);
\draw (11.375,1.25) node[anchor=north]{\scriptsize Through};
\lcryoswitch{10.25}{1.5}
\draw (10.5,1.8) -- (12,1.8);
\draw (11.375,2.05) node[anchor=south]{\scriptsize TWPA};
\rcryoswitch{12}{1.5}
\draw (12.75,0.85) node{\scriptsize Cryogenic} node[anchor=north]{\scriptsize switch} node[anchor=south]{\(\Lsh\)};

\draw (1.2,1.5) node[anchor=east]{\scriptsize Signal in 2} -- (10.25,1.5);
\simptermuddircoup{8.5}{1.5}
\amplifier{11.25}{1.8}
\disolator{6}{1.5}

\draw (1.2,2.45) node[anchor=east]{\scriptsize Qubit 0 control in} -- (5.05,2.45) -- (5.05,1.75);
\draw (1.2,2) node[anchor=east]{\scriptsize Qubit 1 control in} -- (4.85,2) -- (4.85,1.75);
\draw[fill=white, very thick] (4.7,1.25) rectangle(5.2,1.75);
\draw (4.95,1.25) node[anchor=north]{\scriptsize Qubits} node[anchor=south]{\(Q\)};

\coil{11.4}{3.1}
\coilwire[2.5]{11.4}{3.1}
\draw (11.4,2.5+0.3+3.17+0.15) node{\(\vdots\)};

\draw (12.25,1.5) -- (12.75,1.5) -- (12.75,5.5) -- (1.2,5.5) node[anchor=east]{\scriptsize Signal out};
\lramplifier[]{7.5}{5.5}
\draw (7.25,6) node{\scriptsize HEMT};
\disolatorup{12.75}{2.5}

\hattenuator[\sim-60\dB]{3.5}{3.5}
\hattenuator[\sim-70\dB]{3.5}{1.5}
\hattenuator[\sim-70\dB]{3.15}{2}
\hattenuator[\sim-53\dB]{4}{2.45}

\draw (2,1.5) node{\(>\)};
\draw (2,2) node{\(>\)};
\draw (2,2.45) node{\(>\)};
\draw (2,3.5) node{\(>\)};
\draw (2,5.5) node{\(<\)};\end{tikzpicture}
    \caption{
        \textbf{Experimental setup.}
    }
    \label{fig:Setup}
\end{figure}
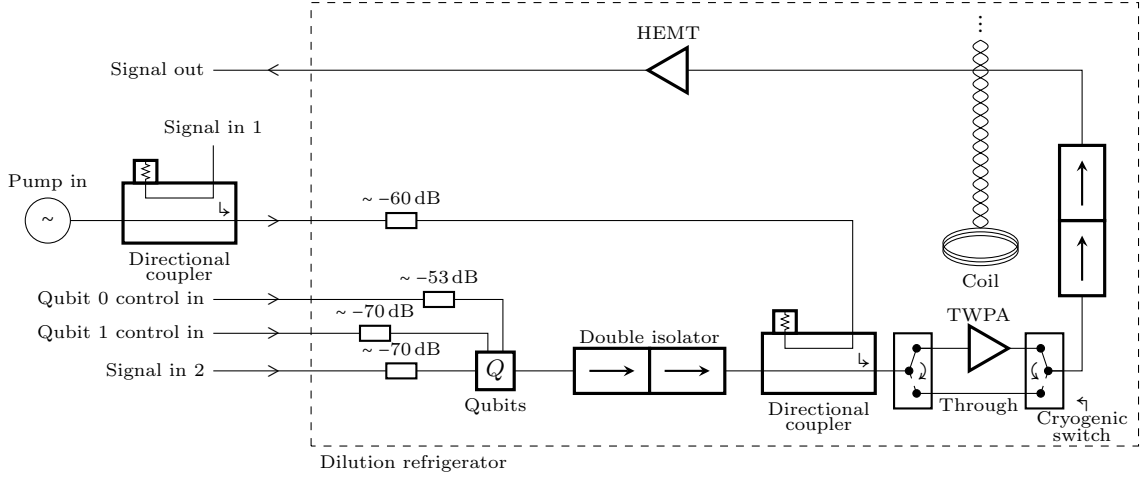

The setup is presented in \Cref{fig:Setup}.
The HEMT amplifier is of the model LNF-LNC0.3\_12A from Low noise factory.
The double isolators are of the model LNF-ISISC4\_12A from Low noise factory.
The room temperature directional coupler is of the model C13-0126 from Marki microwave.
The cryogenic directional coupler is of the model S/N 2C200311002 from Zysentech.

The TWPA is placed between two cryogenic switches, in parallel with a `through' cable.
When measuring gain, a VNA is connected to `Signal in 1' and `Signal out'.
When measuring the \(\Delta\)SNR, the probe signal generator is connected to 
`Signal in 1' and the spectrum analyzer to `Signal out'.
When measuring on our qubit chip,
the qubit readout pulses are applied to the `Signal in 2' port
while the qubit control pulses are applied to the `Qubit \(i\) control in'.

\onecolumngrid

\twocolumngrid

\onecolumngrid
\clearpage
\twocolumngrid

\section{Added noise by a lossy TWPA}
\label{app:Noise}

Our model of the added noise photons by a lossy TWPA, is based on two essential equations.

The first equation is the fundamental theorem for phase-insensitive linear amplifiers \cite{Caves1982},
which describes the equation for the number of added noise photons \(A_i\)
for a phase-preserving amplification process with gain \(G_i\),
\begin{equation}
    A_i \geq \frac{\abs{1 - 1/G_i}}{2}.
\label{eq:AmpNoise}
\end{equation}
Note that the equation allows for both gains, \(G_i>1\), and for attenuations, \(G_i<1\).

The second equation is Friis' formula \cite{Kraus1966},
which describes the total number of added noise photons \(A_\text{tot}\) of an amplifier chain with \(N\) amplifiers,
\begin{equation}
    A_\text{tot} = \sum_{i=1}^N \frac{A_i}{\prod_{j=1}^{i-1} G_j} = A_1 + \frac{A_2}{G_1} + \frac{A_3}{G_1G_2} + ...,
\label{eq:Friis}
\end{equation}
where \(A_i\) is the number of added noise photons of the \(i\):th amplifier
and \(G_j\) is the gain of the \(j\):th amplifier.
Note that the equation allows for both gains, \(G_j>1\), and for attenuations, \(G_j<1\).

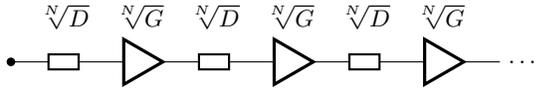
\begin{figure}[H]
    \centering
    \tikzset{external/export next=false} 
\begin{tikzpicture}\draw[fill] (-1.5,0) circle(0.05) -- (5,0) node[anchor=west]{\(\dots\)};
\foreach\x in {-1,1,3} {
    \amplifier{\x+1}0
    \draw (\x+1.2,0.6) node[anchor=center]{\(\sqrt[N]G\)};
    \hattenuator[]{\x}0
    \draw (\x+0.2,0.6) node[anchor=center]{\(\sqrt[N]D\)};
}\end{tikzpicture}
    \caption{
        \textbf{A lossy TWPA.}
        The TWPA consists of \(N\) segments,
        each with an attenuator with the damping factor \(\sqrt[N]D\)
        and the gain factor \(\sqrt[N]G\).
    }
    \label{fig:NoiseAdvanced}
\end{figure}

Now, assume that the TWPA has the total gain \(G\) and the total damping \(D\),
where \(D=1\) means no loss and \(D=0\) means infinite loss,
and that the losses and the gain are evenly distributed.
Then, dividing the TWPA into \(N\) segments,
each segment has the gain \(\sqrt[N]G \geq 1\) and the damping \(\sqrt[N]D \leq 1\),
see \Cref{fig:NoiseAdvanced}.

Inserting \Cref{eq:AmpNoise} into \Cref{eq:Friis},
the total noise added by a TWPA with \(N\) segments can be shown to be
\begin{equation}
\begin{aligned}
    A_{\text{TWPA},N} &= \frac{2\sqrt[N]G - \sqrt[N]{GD} - 1}{2\sqrt[N]{GD}} \sum_{i=0}^{N-1} (GD)^{-i/N} \\ 
    &= \frac{2\sqrt[N]G - \sqrt[N]{GD} - 1}{2\sqrt[N]{GD}} \frac{1 - \frac1{GD}}{1 - \frac1{\sqrt[N]{GD}}} \\
    &= \underbrace{\frac{2\sqrt[N]G - \sqrt[N]{GD} - 1}{\sqrt[N]{GD} - 1}}_\star \frac{GD - 1}{2GD}.
\end{aligned}
\end{equation}
To calculate the continuous limit, \(N\to\infty\), we apply L'Hôpital's rule on the first fraction,
\begin{equation}
\begin{aligned}
    \lim_{N\to\infty} \star &= \lim_{N\to\infty} \frac{2\sqrt[N]G - \sqrt[N]{GD} - 1}{\sqrt[N]{GD} - 1} \\
    &= \lim_{N\to\infty} \frac{-2N^{-2}\ln(G) + N^{-2}\ln(GD)}{-N^{-2}\ln(GD)} \\
    &= \frac{\ln(G/D)}{\ln(GD)}.
\end{aligned}
\end{equation}
The total number of noise photons added is thus
\begin{equation}
    A_\text{TWPA} = \frac{GD - 1}{2GD} \frac{\pln{G/D}}{\pln{GD}}.
\end{equation}
We see that it is not the gain on its own that matters, nor the loss on its own,
but the relation between the gain and the loss.
For a lossless TWPA, \(D=1\), we retrieve the added noise of a lossless amplifier in \Cref{eq:AmpNoise}.
For a gainless TWPA, \(G=1\), we retrieve the added noise of an attenuator in \Cref{eq:AmpNoise}.



\begin{thebibliography}{0}%
\makeatletter
\providecommand \@ifxundefined [1]{%
 \@ifx{#1\undefined}
}%
\providecommand \@ifnum [1]{%
 \ifnum #1\expandafter \@firstoftwo
 \else \expandafter \@secondoftwo
 \fi
}%
\providecommand \@ifx [1]{%
 \ifx #1\expandafter \@firstoftwo
 \else \expandafter \@secondoftwo
 \fi
}%
\providecommand \natexlab [1]{#1}%
\providecommand \enquote  [1]{``#1''}%
\providecommand \bibnamefont  [1]{#1}%
\providecommand \bibfnamefont [1]{#1}%
\providecommand \citenamefont [1]{#1}%
\providecommand \href@noop [0]{\@secondoftwo}%
\providecommand \href [0]{\begingroup \@sanitize@url \@href}%
\providecommand \@href[1]{\@@startlink{#1}\@@href}%
\providecommand \@@href[1]{\endgroup#1\@@endlink}%
\providecommand \@sanitize@url [0]{\catcode `\\12\catcode `\$12\catcode `\&12\catcode `\#12\catcode `\^12\catcode `\_12\catcode `\%12\relax}%
\providecommand \@@startlink[1]{}%
\providecommand \@@endlink[0]{}%
\providecommand \url  [0]{\begingroup\@sanitize@url \@url }%
\providecommand \@url [1]{\endgroup\@href {#1}{\urlprefix }}%
\providecommand \urlprefix  [0]{URL }%
\providecommand \Eprint [0]{\href }%
\providecommand \doibase [0]{http://dx.doi.org/}%
\providecommand \selectlanguage [0]{\@gobble}%
\providecommand \bibinfo  [0]{\@secondoftwo}%
\providecommand \bibfield  [0]{\@secondoftwo}%
\providecommand \translation [1]{[#1]}%
\providecommand \BibitemOpen [0]{}%
\providecommand \bibitemStop [0]{}%
\providecommand \bibitemNoStop [0]{.\EOS\space}%
\providecommand \EOS [0]{\spacefactor3000\relax}%
\providecommand \BibitemShut  [1]{\csname bibitem#1\endcsname}%
\let\auto@bib@innerbib\@empty
\end{thebibliography}%


\begin{thebibliography}{999}
\bibitem{Caves1982}
C. M. Caves, Quantum limits on noise in linear amplifiers, Phys. Rev. D \textbf{26}, 1817 (1982).

\bibitem{Yamamoto2008}
T. Yamamoto, K. Inomata, M. Watanabe, K. Matsuba, T. Miyazaki, W. D. Oliver, Y. Nakamura, and J. S. Tsai,
Flux-driven Josephson parametric amplifier,
Appl. Phys. Lett., \textbf{93}, 042510 (2008).

\bibitem{Bergeal2010}
N. Bergeal, F. Schackert, M. Metcalfe, R. Vijay, V. E. Manucharyan, L. Frunzio, D. E. Prober, R. J. Schoelkopf, S. M. Girvin and M. H. Devoret,
Phase-preserving amplification near the quantum limit with a Josephson ring modulator,
Nature \textbf{465}, 64 (2010).

\bibitem{Roch2012}
N. Roch, E. Flurin, F. Nguyen, P. Morfin, P. Campagne-Ibarcq, M. H. Devoret, and B. Huard,
Widely Tunable, Nondegenerate Three-Wave Mixing Microwave Device Operating near the Quantum Limit,
Phys. Rev. Lett. \textbf{108}, 147701 (2012).

\bibitem{Roy2016}
A. Roy and M. Devoret, Introduction to parametric amplification
of quantum signals with Josephson circuits, Comptes Rendus
Phys. \textbf{17}, 740 (2016).

\bibitem{AumentadoRev}
J. Aumentado,
Superconducting Parametric Amplifiers (Review), 
IEEE Microwave magazine, \textbf{21}, 45-59 (2020).

\bibitem{Girvin2008}
R. J. Schoelkopf and S. M. Girvin, 
Wiring up quantum systems, 
Nature \textbf{451}, 664 (2008).

\bibitem{Kanter1971}
H. Kanter, A. H. Silver,
Self-pumped  Josephson  parametric amplification,
Appl. Phys. Lett., \textbf{19}, 515-517 (1971).

\bibitem{Feldman1975}
M. J. Feldman, P. T. Parrish, R. Y. Chiao,
Parametric amplification by unbiased Josephson junctions,
J. Appl. Phys., \textbf{46}, 4031-4042 (1975).

\bibitem{Yurke1988}
B. Yurke, P. G. Kaminsky, R. E. Miller, E. A. Whittaker, A. D. Smith, A. H. Silver, and R. W. Simon,
Observation of 4.2-K equilibrium-noise squeezing via a Josephson-parametric amplifier,
Phys. Rev. Lett., \textbf{60}, 764-767 (1988).

\bibitem{Yurke1989}
B. Yurke, L. R. Corruccini, P. G. Kaminsky, L. W. Rupp, A. D. Smith, A. H. Silver, R. W. Simon, and E. A. Whittaker,
Observation  of  parametric  amplification  and  deamplification  in  a  Josephson  parametric  amplifier,
Phys. Rev. A, \textbf{39}, 2519-2533 (1989).

\bibitem{Olsson1988}
H. K. Olsson,  T. Claeson,
Low-noise Josephson parametric amplification and oscillations at 9 GHz,
J. Appl. Phys., \textbf{64}, 5234–5243 (1988).

\bibitem{Beltran2008}
M. A. Castellanos-Beltran, K. D. Irwin, G. C. Hilton, L. R. Vale, K. W. Lehnert,
Amplification and squeezing of quantum noise with a tunable Josephson metamaterial
Nature Phys. \textbf{4}, 929 (2008).

\bibitem{Roy2015}
T. Roy, S. Kundu, M. Chand, A. M. Vadiraj, A. Ranadive, N. Nehra, M. P. Patankar, J. Aumentado, A. A. Clerk, and R. Vijay, 
Broadband parametric amplification with impedance engineering: Beyond the gain-bandwidth product
Appl. Phys. Lett. \textbf{107}, 262601 (2015).

\bibitem{Simoen2015}
M. Simoen, C. W. S. Chang, P. Krantz, J. Bylander, W. Wustmann, V. Shumeiko, P. Delsing, C. M. Wilson,
Characterization of a multimode coplanar waveguide parametric amplifier,
J. Appl. Phys. \textbf{118}, 154501 (2015).

\bibitem{Cullen1958}
A. L. Cullen, 
A Travelling-Wave Parametric Amplifier,
Nature (London) \textbf{181}, 332 (1958).

\bibitem{Suhl1958}
P. K. Tien and H. Suhl, 
A Traveling-Wave Ferromagnetic Amplifier,
Proc. lnst. Radio Engrs. \textbf{46}, 700 (1958).

\bibitem{Tien1958} 
P. K. Tien,
Parametric amplification and frequency mixing in propagating circuits,
J. Appl. Phys. \textbf{29}, 1347 (1958).

\bibitem{LeDuc2012}
Byeong Ho Eom, Peter K. Day, Henry G. LeDuc, and Jonas Zmuidzinas,
A wideband, low-noise superconducting amplifier with high dynamic range,
Nature Phys. \textbf{8}, 623 (2012).

\bibitem{Pappas2014}
C. Bockstiegel, J. Gao, M. R. Vissers, M. Sandberg, S. Chaudhuri, A. Sanders, L.R. Vale, K.D. Irwin, and D. P. Pappas,
Development of a Broadband NbTiN TravelingWaveParametric Amplifier for MKID Readout,
J. Low Temp. Phys. \textbf{176}, 476 (2014). 

\bibitem{Pappas2016}
M. R. Vissers, R. P. Erickson, H.-S. Ku, Leila Vale, Xian Wu, G. C. Hilton, and D. P. Pappas,
Low-noise kinetic inductance traveling-wave amplifier using three-wave mixing,
Appl. Phys. Lett. \textbf{108}, 012601 (2016).

\bibitem{Gao2017}
S. Chaudhuri, D. Li, K. D. Irwin, C. Bockstiegel, J. Hubmayr, J. N. Ullom, M. R. Vissers, and J. Gao,
Broadband parametric amplifiers based on nonlinear kinetic inductance artificial transmission lines,
Appl. Phys. Lett. \textbf{110}, 152601 (2017).

\bibitem{Erickson2017}
R. P. Erickson and D. P. Pappas,
Theory of multiwave mixing within the superconducting kinetic-inductance traveling-wave amplifier,
Phys. Rev. B \textbf{95}, 104506 (2017).

\bibitem{Katz2020}
S. Goldstein, N. Kirsh, E. Svetitsky, Y. Zamir, O. Hachmo, C.E. Mazzotti de Oliveira, and N. Katz,
Four wave-mixing in a microstrip kinetic inductance travelling wave parametric amplifier,
Appl. Phys. Lett. \textbf{116}, 152602 (2020).

\bibitem{GaoPRXQ2021}
M. Malnou, M. R. Vissers, J. D. Wheeler,  J. Aumentado, J. Hubmayr, J. N. Ullom, and J. Gao,
Three-Wave Mixing Kinetic Inductance Traveling-Wave Amplifier with Near-Quantum-Limited Noise Performance,
PRX Quantum \textbf{2}, 010302 (2021).

\bibitem{Duti2021}
D. J. Parker, M. Savytskyi, W. Vine, A. Laucht, T. Duty, A. Morello, A.L. Grimsmo, and J.J. Pla,
Degenerate Parametric Amplification via Three-Wave Mixing Using Kinetic Inductance,
Phys. Rev. Appl. \textbf{17} 034064 (2022).

\bibitem{Aalto2021}
M.R. Perelshtein, K.V. Petrovnin, V. Vesterinen, S. Hamedani Raja, I. Lilja, M. Will, A. Savin, S. Simbierowicz, R.N. Jabdaraghi, J.S. Lehtinen, L. Gr\"onberg, J. Hassel, M.P. Prunnila, J. Govenius, G.S. Paraoanu, and P.J. Hakonen,
Broadband Continuous-Variable Entanglement Generation Using a Kerr-Free Josephson Metamaterial,
Phys. Rev. Appl. \textbf{18}, 024063 (2022).

\bibitem{Siddiqi2013}
O. Yaakobi, L. Friedland, C. Macklin, and I. Siddiqi,
Parametric amplification in Josephson junction embedded transmission lines,
Phys. Rev. B \textbf{87}, 144301 (2013).

\bibitem{Obrien2014} 
K. O'Brien, C. Macklin, I. Siddiqi, and Xiang Zhang,
Resonant Phase Matching of Josephson Junction Traveling Wave Parametric Amplifiers,
Phys. Rev. Lett. \textbf{113}, 157001 (2014).

\bibitem{Macklin2015}
C. Macklin, K. O'Brien, D. Hover, M. E. Schwartz, V. Bolkhovsky, X. Zhang, W. D. Oliver, I. Siddiqi,
A near-quantum-limited Josephson traveling wave parametric amplifier,
Science \textbf{350}, 325 (2015).

\bibitem{Martinis2015} 
T. C. White, J. Y. Mutus, I.-C. Hoi, R. Barends, B. Campbell, Yu Chen, Z. Chen, B. Chiaro, A. Dunsworth, E. Jeffrey, J. Kelly, A. Megrant, C. Neill, P. J. J. O'Malley, P. Roushan, D. Sank, A. Vainsencher, J. Wenner, S. Chaudhuri, J. Gao and John M. Martinis,
Traveling wave parametric amplifier with Josephson junctions using minimal resonator phase matching,
Appl. Phys. Lett. \textbf{ 106}, 242601 (2015).

\bibitem{Bell2015}
M.T. Bell and A. Samolov,
Traveling-Wave Parametric Amplifier Based on a Chain of Coupled Asymmetric SQUIDs,
Phys. Rev. Applied \textbf{4}, 024014 (2015).

\bibitem{Planat2020}
L. Planat, A. Ranadive, R. Dassonneville, J. Puertas Martínez, S. Léger, C. Naud, O. Buisson, W. Hasch-Guichard, D.M. Basko and N. Roch,
Photonic-Crystal Josephson Traveling-Wave Parametric Amplifier
Phys. Rev. X \textbf{10}, 021021 (2020).

\bibitem{Ranadive2022}
A. Ranadive, M. Esposito, L. Planat, E. Bonet, C. Naud, O. Buisson, W. Guichard, and N. Roch,
Kerr reversal in Josephson meta-material and traveling wave parametric amplification,
Nature Commun \textbf{13}, 1737 (2022).

\bibitem{Zorin2016}
A. B. Zorin,
Josephson Traveling-Wave Parametric Amplifier with Three-Wave Mixing,
Phys. Rev. Appl., \textbf{6}, 034006 (2016).

\bibitem{Zorin2019}
A. B. Zorin,
Flux-Driven Josephson Traveling-Wave Parametric Amplifier,
Phys. Rev. Appl., \textbf{12}, 044051 (2019).

\bibitem{Sivak2019}
V. V. Sivak, N. E. Frattini, V. R. Joshi, A. Lingenfelter, S. Shankar, and M. H. Devoret,
Kerr-Free Three-Wave Mixing in Superconducting Quantum Circuits,
Phys. Rev. Appl., \textbf{11}, 054060 (2019).

\bibitem{Mukhanov2019}
A. Miano and O. A. Mukhanov,
Symmetric Traveling Wave Parametric Amplifier,
IEEE Trans. Appl. Supercond. \textbf{29}, 1501706 (2019).

\bibitem{MyTWPA3WM}
H. Renberg Nilsson, A. Fadavi Roudsari, D. Shiri, P. Delsing, and V. Shumeiko, 
High-Gain Traveling-Wave Parametric Amplifier Based on Three-Wave Mixing,
Phys. Rev. Appl., \textbf{19}, 044056 (2023).

\bibitem{AnitasArticle}
A. Fadavi Roudsari, D. Shiri, H. Renberg Nilsson, G. Tancredi, A. Osman, I. Svensson, M. Kudra, M. Rommel, J. Bylander, V. Shumeiko, and P. Delsing,
Three-wave mixing traveling-wave parametric amplifier with periodic variation of the circuit parameters,
Appl. Phys. Lett., \textbf{122}, 052601 (2023).

\bibitem{Gaydamachenko2022}
Victor Gaydamachenko, Christoph Kissling, Ralf Dolata, Alexander B. Zorin,
Numerical analysis of a three-wave-mixing Josephson traveling-wave parametric amplifier with engineered dispersion loadings,
J. Appl. Phys. \textbf{21}, 154401 (2022).

\bibitem{Frattini2017}
N. E. Frattini, U. Vool, S. Shankar, A. Narla, K. M. Sliwa, and M. H. Devoret,
3-wave mixing Josephson dipole element,
Appl. Phys. Lett. \textbf{110}, 222603 (2017).

\bibitem{RenbergNilsson2024}
H. Renberg Nilsson, D. Shiri, R. Rehammar, A. Fadavi Roudsari and P. Delsing,
Peripheral circuits for ideal performance of a traveling-wave parametric amplifier,
Phys. Rev. Appl. \textbf{21}, 064062 (2024).

\bibitem{Kow2024}
C. Kow, V. Podolskiy, A. Kamal,
Self phase-matched broadband amplification with a left-handed Josephson transmission line,
arXiv:2201.04660 (2024).

\bibitem{MatthaeiYoungJones}
G. L. Matthaei, L. Young and E. M. T. Jones,
\textit{Microwave filters, impedance-matching networks, and coupling structures},
(McGraw-Hill Book Company, Inc., Norwood 1964).

\bibitem{HRNthesis}
H. Renberg Nilsson,
Superconducting lumped-element travelling-wave parametric amplifiers,
PhD thesis, Chalmers University of Technology (2024).

\bibitem{Khor2003}
K. A. Khor, K. H. Cheng, L. G. Yu and F. Boey,
Thermal conductivity and dielectric constant of spark plasma sintered aluminum nitride,
Materials Science and Engineering: A, \textbf{347}, 300-305 (2003).

\bibitem{Kosen2022}
Kosen, S. et al.,
Building blocks of a flip-chip integrated superconducting quantum processor,
Quantum Sci. Technol. \textbf{7}, 035018 (2022).

\bibitem{Koch2007}
Koch, J. et al.,
Charge-insensitive qubit design derived from the cooper pair box,
Phys. Rev. A \textbf{76}, 042319 (2007).

\bibitem{Magesan2015}
Magesan, E., Gambetta, J. M., Córcoles, A. D. \& Chow,
J. M. , Machine learning for discriminating quantum measurement trajectories and improving readout,
Phys. Rev. Lett. \textbf{114}, 200501 (2015).

\bibitem{Chen2023}
Chen, L., Li, HX., Lu, Y. et al.,
Transmon qubit readout fidelity at the threshold for quantum error correction without a quantum-limited amplifier.,
npj Quantum Inf \textbf{9}, 26 (2023).

\bibitem{Walter2017}
Walter, T. et al.,
Rapid high-fidelity single-shot dispersive readout of superconducting qubits,
Phys. Rev. Appl. \textbf{7}, 054020 (2017).

\bibitem{PatentOne}
Swedish patent P440912PC00.

\bibitem{Kraus1966}
J. D. Kraus,
Radio astronomy,
McGraw-Hill (1966).

\end{thebibliography}
\end{document}